# Methylator: A Modular Framework for DNA Methylation Analysis in Mammals and Plants Using Galaxy

Jonas Bucher[1,2], Ueli Grossniklaus[1,2,3], Deepak Kumar Tanwar[1,2,4,5]*

[1]Department of Plant and Microbiology, University of Zurich, Zollkerstrasse 107, 8008 Zurich, Switzerland
[2]University Research Priority Program "Evolution in Action", University of Zurich, Zollikerstrasse 107, 8008 Zurich, Switzerland
[3]Zurich-Basel Plant Science Center, University of Zurich, ETH Zurich & University of Basel, Tannenstrasse 1, 8092 Zurich, Switzerland
[4]Present Address: Swiss Institute of Bioinformatics, Winterthurerstrasse 190, 8057 Zurich, Switzerland.
[5]Present Address: Department of Quantitative Biomedicine, University of Zurich, Winterthurerstrasse 190, 8057 Zurich, Switzerland.

*Corresponding author: Deepak Kumar Tanwar, University of Zurich and Swiss Institute of Bioinformatics, Winterthurerstrasse 190, 8057 Zurich, Switzerland.
E-mail: deepak.tanwar@uzh.ch

## Abstract

### Summary

DNA cytosine methylation is a critical epigenetic mark regulating gene expression and thus playing an important role in development and differentiation across eukaryotes. Existing tools for high-throughput methylation analysis often lack cross-species flexibility or require command-line expertise. We present Methylator, a novel, end-to-end DNA methylation analysis framework integrated into the Galaxy platform, enabling accessible DNA methylation analysis for mammals and plants. Methylator supports analyses from data obtained using diverse protocols like WGBS, RRBS, and PBAT and handles all contexts of DNA methylation (CpG, CHG, and CHH). The Methylator framework includes quality control, alignment, methylation calling, differential analysis, and functional analysis through reproducible, user-friendly workflows. Its unique Dirty-Harry alignment method enhances mapping efficiency, while a Shiny-based interface allows for interactive, publication-ready visualizations. 𝙼𝚎𝚝𝚑𝚢𝚕𝚊𝚝𝚘𝚛 is freely available, offering researchers a versatile, user-friendly solution for epigenomic studies.

Availability and Implementation

𝙼𝚎𝚝𝚑𝚢𝚕𝚊𝚝𝚘𝚛 is built on Galaxy version 24.1, and is freely available as a Docker image with open-source code on GitHub under the MIT license: https://github.com/urppeia/methylator-galaxy.

---

# Introduction

DNA cytosine methylation is an evolutionarily conserved modification with key roles in gene regulation, epigenomic stability, and phenotypic plasticity. In mammals, methylation is predominantly at CpG dinucleotides near promoters, influencing gene expression and being involved in genomic imprinting and X-chromosome inactivation. In plants, such as the model *Arabidopsis thaliana*, methylation extends to the CHG and CHH contexts (H referring to A, T or C). While CpG methylation is primarily found in gene bodies, transposable elements and repetitive regions are heavily methylated in all contexts (Law & Jacobsen, 2010; Zhang et al., 2018). Advances in bisulfite sequencing methods, such as whole-genomic bisulfite sequencing (WGBS), reduced representation bisulfite sequencing (RRBS), and post-bisulfite adaptor tagging (PBAT), enable genome-wide, base-resolution profiling of DNA methylation. However, bisulfite treatment introduces specific challenges, including reduced sequence complexity, read fragmentation, and biased coverage. These issues complicate alignment and downstream interpretation, particularly for researchers working with non-model organisms or low-quality input. Furthermore, the computational tools needed to process such data are often fragmented across platforms, require scripting knowledge, or are poorly documented.

## Implementation

To address these above-mentioned challenges, we developed 𝙼𝚎𝚝𝚑𝚢𝚕𝚊𝚝𝚘𝚛, a comprehensive framework built entirely within Galaxy, a graphical workflow system for bioinformatic analyses (Afgan et al., 2018). The 𝙼𝚎𝚝𝚑𝚢𝚕𝚊𝚝𝚘𝚛 framework was developed using Planemo, a command-line toolkit for building Galaxy tools. 𝙼𝚎𝚝𝚑𝚢𝚕𝚊𝚝𝚘𝚛 enables researchers to run complete methylation pipelines — from raw reads to biological interpretation — without requiring programming skills or external tools. All workflows are automated, version-controlled, and graphically configurable, ensuring ease of use and reproducibility.

# Workflow Overview

## Quality control

`Methylator` processes raw data through a streamlined pipeline, Figure 1, Supplementary Figure 1 and 2. *FastQC* assesses quality, followed by *Trim Galore* (Krueger, 2023) for adapter and quality trimming. *Clumpify* deduplicates reads, removing optical, PCR, and tile-edge duplicates based on platform-specific spatial patterns (Bushnell, 2014). Additionally, *Clumpify* reorders reads to improve compression and reduce alignment time.

Alignment and methylation calling are performed using *Bismark* (Krueger & Andrews, 2011), enhanced with the `--local` option to support the *Dirty Harry* method for alignment. This method improves mapping efficiency by salvaging unmapped paired-end reads through secondary alignment in local single-end mode. For directional libraries, Read 1 is aligned using default parameters and Read 2 with `--pbat`; for non-directional libraries, the roles are reversed (Andrews et al., 2023). After alignment, the *Bismark* methylation extractor is applied separately to all three BAM files (paired-end and two single-end), followed by merging into a unified coverage and cytosine report per sample. This process recovers substantial amounts of methylation data typically lost in standard workflows.

## Differential Methylation Analysis

`Methylator` includes parallel differential analysis at both, the single-cytosine and regional levels:

- ***methylKit*** (Akalin et al., 2012) identifies differentially methylated loci (DMLs) using logistic regression or Fisher's exact test, filtering out low-coverage or excessively covered sites.

- ***dmrseq*** (Korthauer et al., 2019) detects differentially methylated regions (DMRs) using a two-step method that accounts for spatial correlation and controls false discovery rate.

This dual-resolution approach provides granular as well as regulatory-scale insights into DNA methylation dynamics, accommodating both focused and broad hypotheses.

## Functional Annotation and Motif Analysis

To associate methylation changes with gene functions, *rGREAT* is used to map DMRs and DMLs to Gene Ontology (GO) terms, KEGG, and Reactome pathways (McLean et al., 2010; Gu & Hübschmann, 2022). This includes automatic stratification into biological processes, molecular functions, and cellular components. Enrichment results are visualized using dot plots, volcano plots, and association histograms, Supplementary Figure 5.

Motif enrichment is performed using *monaLisa* (Machlab et al., 2022), which identifies enriched DNA sequence motifs in differentially methylated regions. The results are visualized via interactive heatmaps with embedded sequence logos, offering a functional view of regulatory sequence patterns, Supplementary Figure 5.

### Customization and Data Types

Methylator supports multiple sequencing protocols including WGBS, RRBS, and PBAT, as well as cytosine context options (CpG-only or all contexts). Workflow parameters are automatically optimized based on the provided metadata, which includes sequencing platform, reference genome, and grouping variables for differential analysis.

### Exploratory Data Analysis via Shiny

All results from the differential methylation, functional, and motif enrichment analyses are compiled into a single, user-friendly Shiny application, Supplementary Figure 3. This stand-alone web-based application allows users to upload and visually explore their result data. Figures can be easily customized in terms of size, text, labels, and color schemes, including recommended colorblind-friendly options, Supplementary Figure 4. Users can download figures in multiple formats, ensuring convenient integration of the analysis outputs into further research or publications, and closing the loop from raw data to biological insights.

## Acknowledgements

We thank all members of the UG laboratory for insightful discussions. We also gratefully acknowledge Dr. Masaomi Hatakeyama for valuable feedback and helpful discussions.

## Author contributions

Conceptualization: DKT; Funding acquisition: UG, DKT; Methodology: JB, DKT; Software: JB; Project administration: UG; Resources: UG; Supervision: UG, DKT; Writing—original draft: JB; Writing—review & editing: UG, DKT.

## Supplementary data

Supplementary figures and tables are provided with the manuscript.

## Conflict of interest:

None declared.

## Funding

This project was supported by the University Research Priority Program "Evolution in Action" and core funding from the University of Zurich to UG.

## Data availability

No dataset has been generated in this study.

## Code availability

Methylator with Galaxy: https://github.com/urppeia/methylator-galaxy
Shiny application for exploratory data analysis: https://github.com/urppeia/DNAme_shiny
Galaxy tools developed: https://github.com/urppeia/galaxy/tree/dev/tools/my_tools

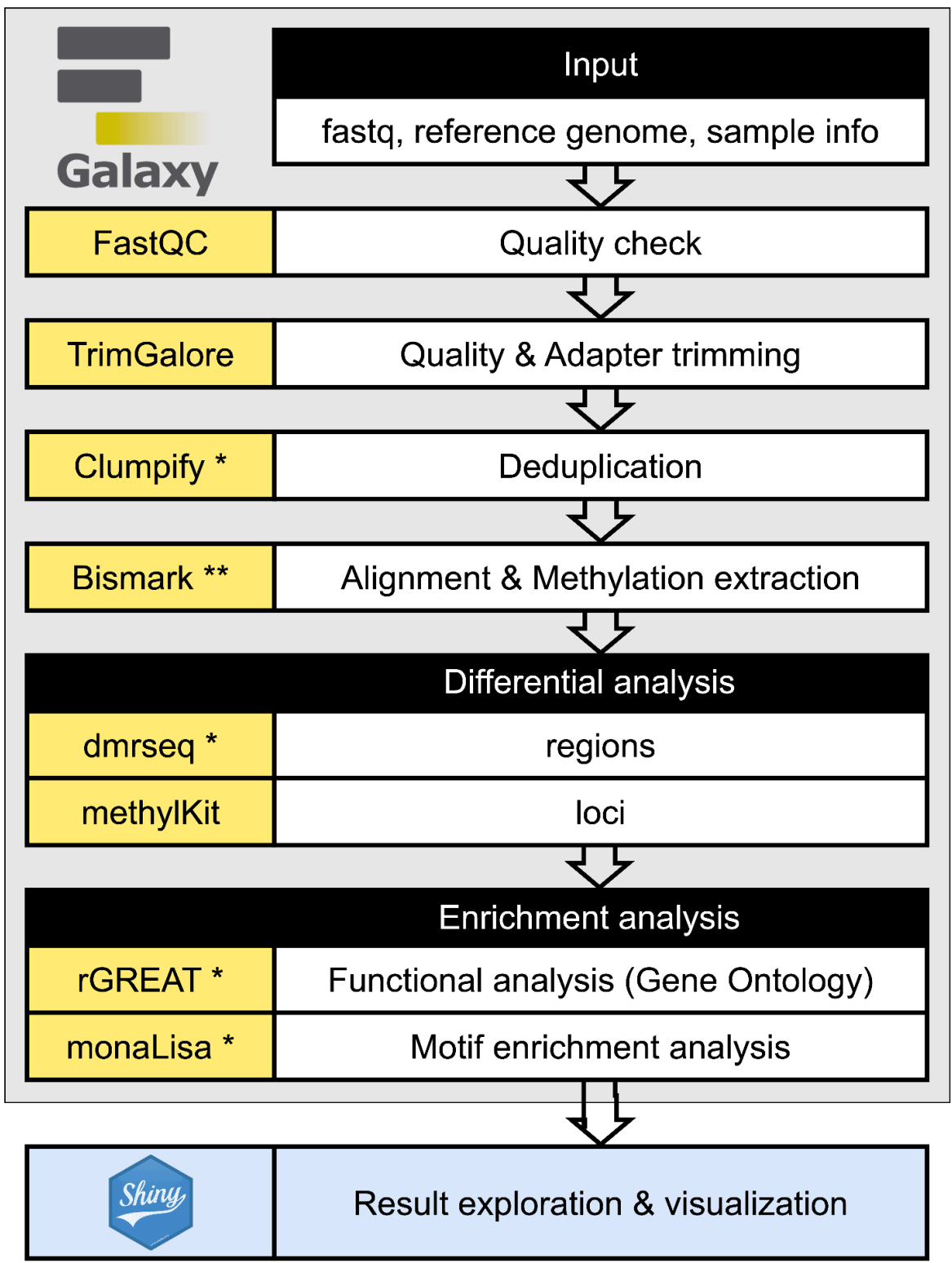

**Figure 1**. Methylator workflow in Galaxy.
* Method previously not implemented in Galaxy
** Method previously partially implemented in Galaxy

# Supplementary Data

## Supplementary Methods

The supplementary methods section provides a comprehensive description of the tools, techniques, and workflows utilized in the Methylator framework, integrated into the Galaxy platform (version 24.1).

## Tools Used in Methylator Framework

The Methylator pipeline leverages a robust set of bioinformatics tools to facilitate end-to-end DNA methylation analysis across mammals and plants:

- **FastQC**: Employed for quality assessment of raw sequencing reads to ensure data integrity prior to processing.

- **Trim Galore**: Used for adapter and quality trimming, optimizing reads for downstream alignment.

- **Clumpify**: Applied for deduplication, removing optical, PCR, and tile-edge duplicates based on platform-specific spatial patterns, and reordering reads to enhance compression.

- **Bismark**: Utilized for alignment and methylation calling, enhanced with the --local option to implement the Dirty Harry method, which improves mapping efficiency by salvaging unmapped paired-end reads through secondary alignment in local single-end mode.

- **methylKit**: Conducts differential methylation analysis at the single-cytosine level (DMLs) using logistic regression or Fisher's exact test, filtering out low-coverage or excessively covered sites.

- **dmrseq**: Performs differential methylation analysis at the regional level (DMRs), accounting for spatial correlation and controlling the false discovery rate with a two-step method.

- **rGREAT**: Facilitates functional annotation by mapping DMRs and DMLs to Gene Ontology (GO) terms, KEGG, and Reactome pathways, with results visualized through dot plots, volcano plots, and association histograms.

- **monaLisa**: Implements motif enrichment analysis, identifying enriched DNA sequence motifs in differentially methylated regions, with results presented via interactive heatmaps and embedded sequence logos.

## Workflow Overview

The Methylator workflow, as illustrated in Figure 1 of the manuscript, processes raw data through a streamlined pipeline within Galaxy. The sequence of steps includes:

- **Quality Control**: FastQC assesses read quality, followed by Trim Galore for adapter and quality trimming. Clumpify deduplicates reads and reorders them to improve compression.

- **Alignment and Methylation Calling**: Bismark, enhanced with the Dirty Harry method, aligns reads and extracts methylation data, recovering substantial amounts typically lost in standard workflows.

- **Differential Methylation Analysis**: Parallel analyses are conducted using methylKit for loci (DMLs) and dmrseq for regions (DMRs), providing dual-resolution insights.

- **Functional and Motif Enrichment Analysis**: rGREAT annotates DMRs and DMLs with GO terms, KEGG, and Reactome pathways, while monaLisa identifies enriched motifs.

- **Exploratory Data Analysis**: A Shiny-based interface compiles results, enabling interactive visualization and customization of figures for publication.

## User Inputs and Parameters

Methylator allows users to define specific parameters within the data analysis pipeline through a graphical user interface (GUI). Input files in fastq format, obtained from high-throughput sequencing (HTS) machines, are supported from various library preparation methods, including whole-genome bisulfite sequencing (WGBS), reduced representation bisulfite sequencing (RRBS), post-bisulfite adapter tagging (PBAT), TET-assisted pyridine borane sequencing (TAPS), anchor-based bisulfite sequencing (ABBS), and enzymatic methyl sequencing (EM-seq). The Methylator framework is suitable for both mammalian and plant species, including polyploids. Utilization of the Methylator pipeline can be carried out in the Galaxy platform, requiring users to provide the reference genome. Users are additionally required to define essential parameters for various aspects of the pipeline, such as the species of their data. However, the most appropriate or tool-default parameters for various analyses have been pre-selected. Moreover, Methylator permits users to opt for exclusive analysis of the CpG context or include the CHG and CHH contexts, with the latter being recommended for plant genome methylation studies.

## Quality and Adapter Trimming

Raw reads in fastq format are initially processed using FastQC (Andrews, 2010) for quality assessment. Subsequently, Trim Galore (Krueger, 2023) is employed for adapter and quality trimming. Trim Galore accommodates specific needs by providing input options that are fine-tuned to ensure appropriate quality and adapter trimming for different types of input data, such as those from PBAT or RRBS libraries.

## Deduplication

Deduplication is performed using Clumpify from the BBMap suite (Bushnell, 2014). Clumpify identifies and removes technical duplicates based on spatial patterns specific to the sequencing platform. These include sister duplicates (originating from complementary strands forming independent clusters), optical duplicates (misidentifying a single cluster as two), tile-edge duplicates (arising numerously on X-patterned flow cells/NextSeq), and clustering duplicates (a single library occupying two neighboring wells during cluster generation, unique to patterned flow cells). Duplicates that do not fall within these categories are assumed to be biological replicates or PCR artifacts. The type of duplicates present and the respective Clumpify options used for different sequencing platforms are summarized in Table S1. In addition to removing duplicates, Clumpify reorders the reads to improve data compression, resulting in approximately 30% smaller files without affecting subsequent analysis.

**Table S1.** Duplicates and Clumpify options for different sequencing platforms.

## Alignment and Methylation Calling

Alignment and methylation calling are performed using Bismark (Krueger and Andrews, 2011), with Bowtie2 as the underlying aligner. In the case of polyploid species, EAGLE-RC (Kuo et al., 2018) is used after alignment to classify reads based on genotype differences between parental genomes and mark the progenitor origin of reads.

For paired-end libraries, the *Dirty Harry* method is optionally applied to improve mapping efficiency. Bisulfite treatment commonly results in decreased sequence complexity and shorter fragments, leading to reduced mapping rates. The *Dirty Harry* method performs paired-end alignment using the --unmapped option, which stores unmapped reads into separate files. These unmapped reads are then aligned separately in local single-end mode. For directional libraries, Read 1 is aligned using default options while Read 2 is mapped using --pbat. For non-directional libraries, these options are reversed. The Bismark methylation extractor is run on the BAM files of the single-end mapped reads, resulting in three separate coverage and cytosine report files per sample (paired-end and two single-end). These are merged in subsequent steps to produce a single coverage file and cytosine report per sample.

## Differential Methylation Analysis

Differential methylation analysis is conducted at both the single-cytosine (loci) and regional levels.

### Loci

Analysis on loci is performed using the R package methylKit (Akalin et al., 2012). Coverage files are read and filtered for each cytosine context separately. Bases with coverage lower than 5 reads (customizable) or having coverage lower than 0.1% or higher than 99.9% are removed to increase significance and eliminate PCR bias. Candidate differentially methylated loci (DML) between two groups are calculated using Fisher's exact test or logistic regression (automatically selected if multiple samples per group are present) via calculateDiffMeth. Significant loci are obtained and split into hypo- and hyper-methylated loci using getMethylDiff with a threshold difference of 25% between groups and a q-value cutoff of 0.01.

### Regions

Regional analysis is carried out using dmrseq (Korthauer et al., 2019). Loci with coverage of at least one per sample are analyzed; others are filtered out. Candidate regions are determined using a default cutoff value of 0.1 (adaptable). Significantly differentially methylated regions (DMR) are selected using a q-value (false discovery rate, FDR) cutoff of 0.05 (adaptable). dmrseq accounts for spatial correlation and controls FDR using the Benjamini-Hochberg method.

## Functional Annotation

Functional analysis associates DML and DMR with Gene Ontology (GO) terms, Kyoto Encyclopedia of Genes and Genomes (KEGG) pathways, and Reactome pathways using the R package rGREAT (Gu and Hübschmann, 2022). Results from DML and DMR analysis are split into different contexts and hypo-/hyper-methylated sets. These are tested against methylated regions using the mode Basal plus extension with default values and compared via a binomial test. Ontology terms are obtained from BioMart. The analysis is performed individually for biological processes, molecular functions, and cellular components. For Arabidopsis thaliana, Reactome and KEGG pathways are additionally included.

## Motif Enrichment Analysis

Motif enrichment is performed using monaLisa (Machlab et al., 2022) to identify enriched DNA sequence motifs in differentially methylated regions. Results from differential analysis of loci and regions are converted to BED files for input. Motif abundance within significantly differentially methylated regions is compared to background regions (full set of differentially methylated regions before q-value filtering) using appropriate parameters, with results visualized via interactive heatmaps and embedded sequence logos.

## Exploratory Data Analysis via Shiny

All results from differential methylation, functional, and motif enrichment analyses are compiled into a single, user-friendly Shiny application. This stand-alone web-based application allows users to upload and visually explore their result data. Figures can be customized in terms of size, text, labels, and color schemes, including colorblind-friendly options. Users can download figures in multiple formats (e.g., PNG, PDF), ensuring convenient integration into further research or publications.

## Comparison with Other Tools

A comparison of features with widely used tools is provided in Table S2.

**Table S2.** Comparison of Methylator with other tools.

# Supplementary Figures Captions

## Supplementary Figure 1

**Galaxy workflow: Preprocessing and alignment.**
Overview of the initial steps in the Methylator Galaxy pipeline, including raw FASTQ input, quality control with FastQC, adapter trimming using Trim Galore, deduplication with Clumpify, and bisulfite alignment via Bismark. The workflow also includes optional single-end rescue for unmapped paired-end reads and methylation extraction to generate coverage and cytosine reports.

## Supplementary Figure 2

**Galaxy workflow: Downstream methylation analysis.**
Post-alignment analysis steps comprising differential methylation detection using methylKit (locus-level) and dmrseq (region-level), functional annotation with rGREAT, and motif enrichment analysis via monaLisa. Outputs are formatted for visualization in the Shiny app.

## Supplementary Figure 3

**Shiny interface: Data upload page.**
Interface for uploading CpG, CHG, and CHH context data. Users can provide outputs from dmrseq, methylKit, rGREAT, and monaLisa, along with sample metadata. The modular design ensures seamless integration of Galaxy outputs into the visualization environment.

## Supplementary Figure 4

**Shiny interface: Settings page.**
Customization options for visualization, including cytosine context selection, methylation status filters (hypo-, hyper-, all), colorblind-friendly palettes (Dark2, Paired, Set2), and manual color assignment for groups and samples to produce consistent, publication-ready figures.

## Supplementary Figure 5

**Example functional enrichment visualization.**
Dot plot generated by rGREAT, summarizing Gene Ontology enrichment for hypo- and hyper-methylated regions. The plot displays term significance, gene counts, and adjusted p-values, facilitating biological interpretation of methylation changes.

Supplementary Figure 6.

**Motif enrichment heatmap.**
Interactive heatmap from monaLisa, highlighting enriched DNA sequence motifs within differentially methylated regions. Embedded sequence logos provide visual representations of motif patterns associated with regulatory elements.

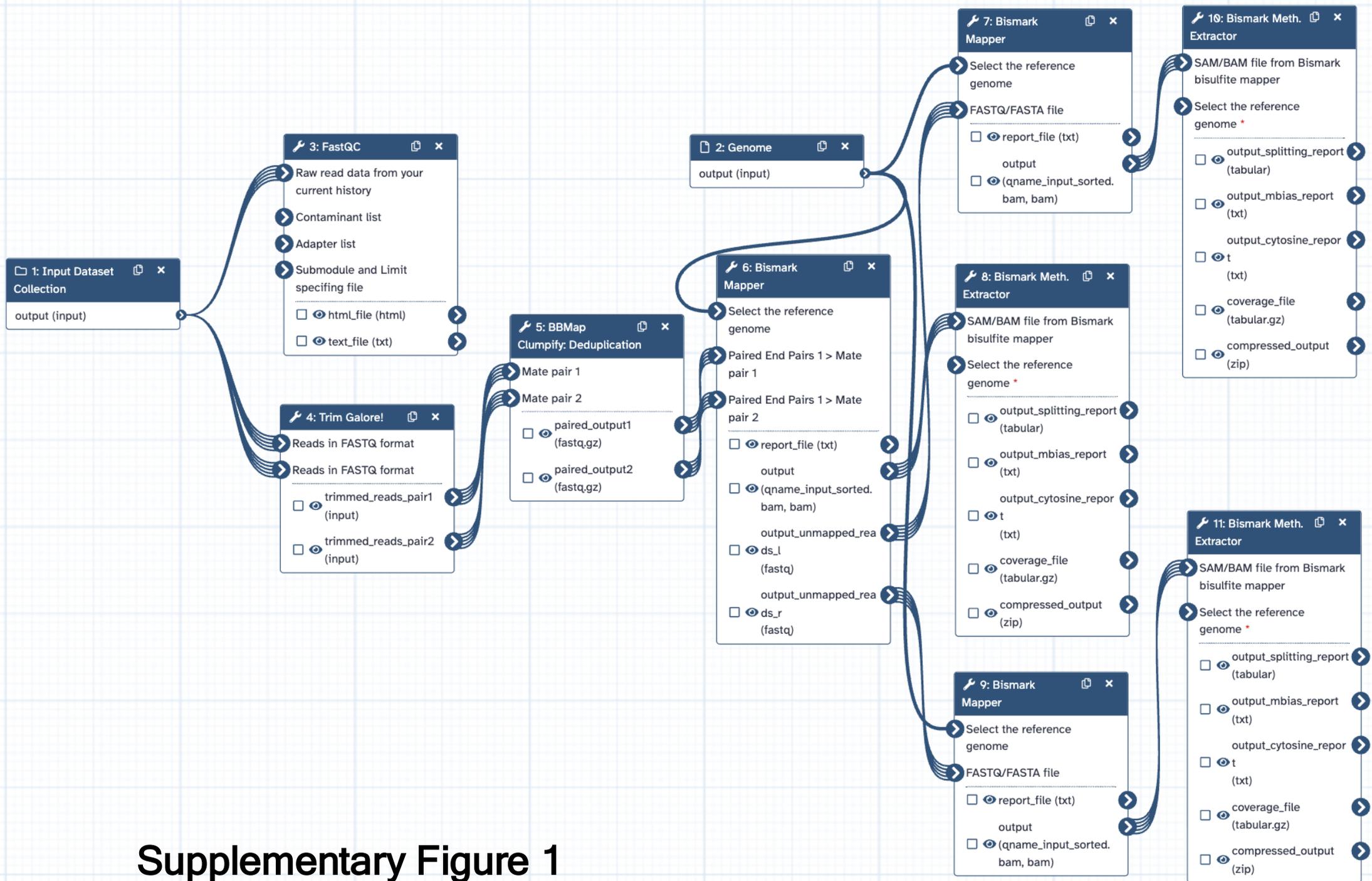

Supplementary Figure 1

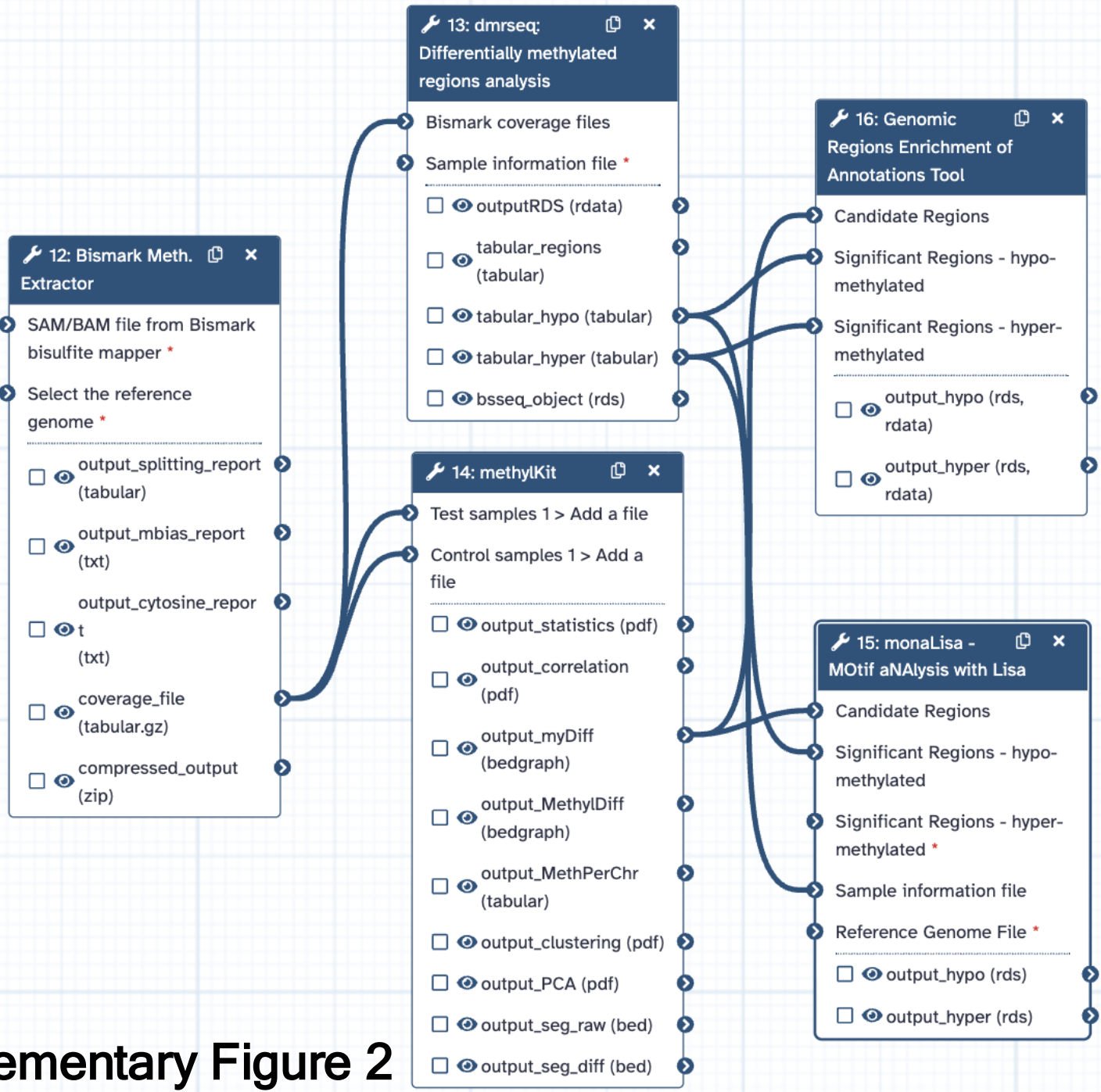

Supplementary Figure 2

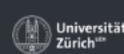

**Supplementary Figure 3**

**Supplementary Figure 4**

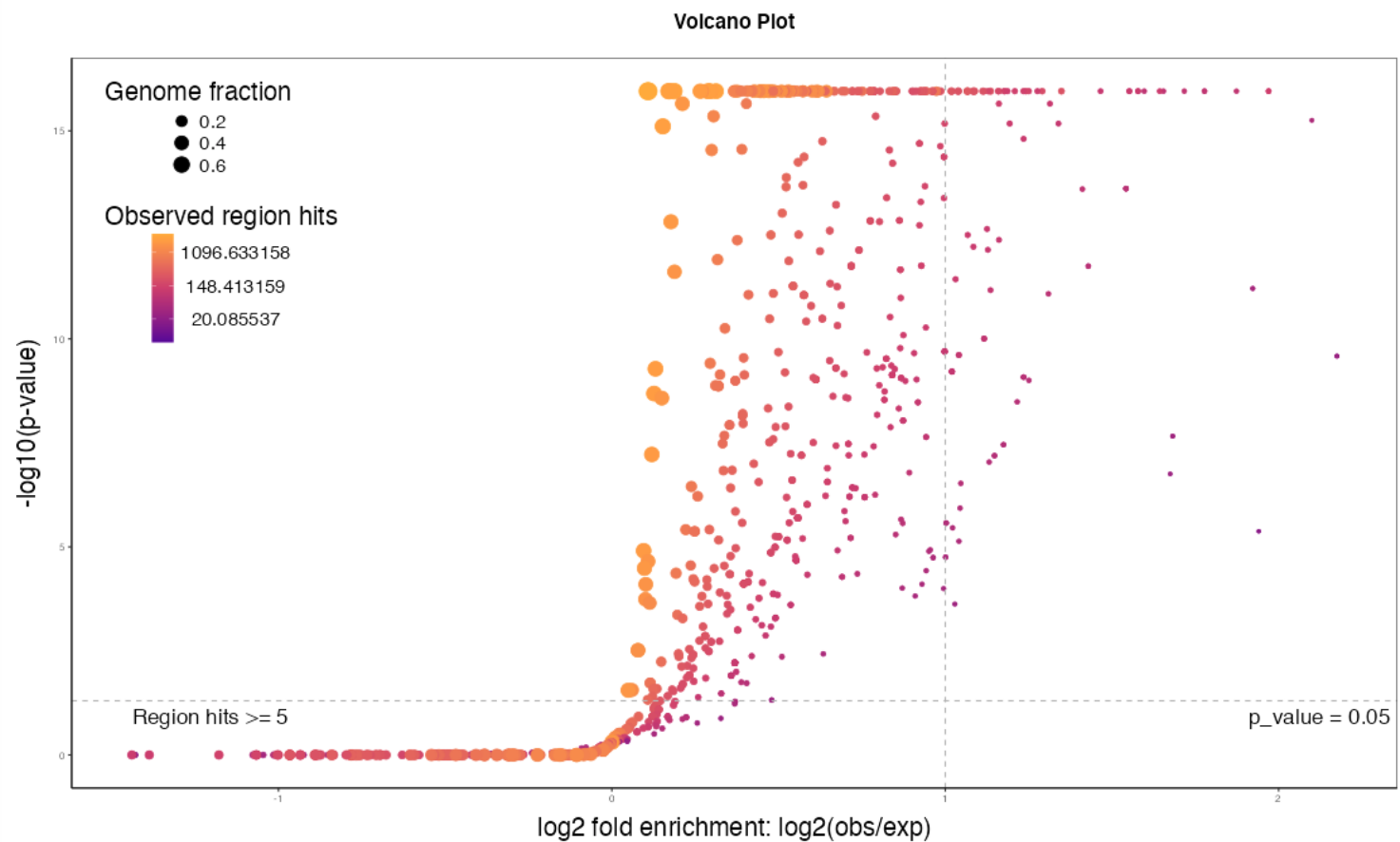

Supplementary Figure 5

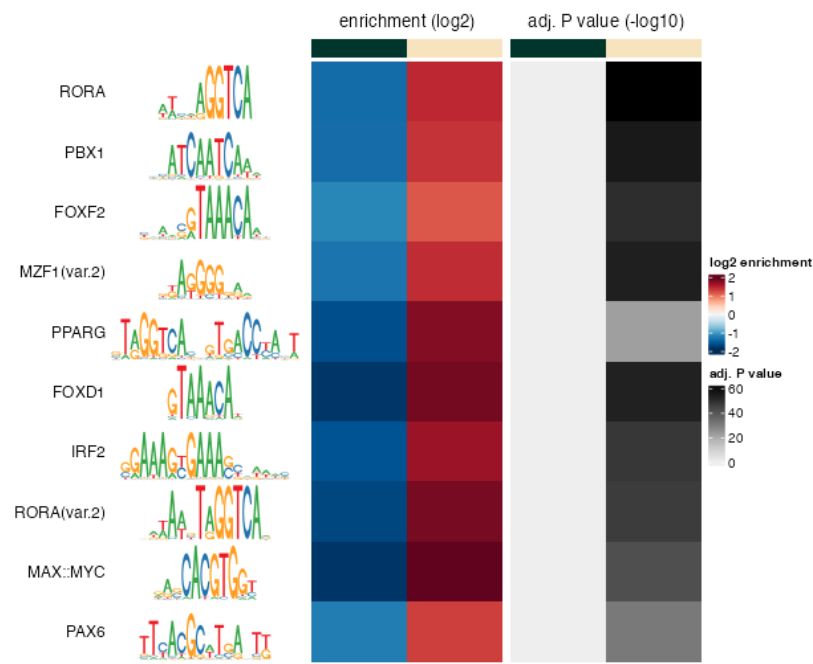

**Supplementary Figure 6**